\documentclass[twocolumn]{aastex631}

\usepackage{graphicx}
\usepackage[none]{hyphenat}
\usepackage{amsmath}
\usepackage{booktabs}
\usepackage{multirow}
\usepackage{txfonts}
\usepackage{enumitem}
\usepackage{algorithmic}
\usepackage{bm}
\usepackage{mathrsfs}
\usepackage{booktabs}  
\usepackage{float}
\usepackage{subfigure}
\DeclareRobustCommand{\VAN}[3]{#2}
\let\VANthebibliography\thebibliography
\def\thebibliography{\DeclareRobustCommand{\VAN}[3]{##3}\VANthebibliography}
\usepackage{color}

\begin{document}

\title{Revealing the spectral properties of magnetic turbulence by synchrotron polarization gradients
}

\author{Jian-Fu Zhang}
\affiliation{Department of Physics, Xiangtan University, Xiangtan, Hunan 411105, People’s Republic of China;\\}\email{jfzhang@xtu.edu.cn}
\affiliation{Key Laboratory of Stars and Interstellar Medium, Xiangtan University, Xiangtan 411105, People’s Republic of China\\}

\author{Zi-Qi Liu}
\affiliation{Department of Physics, Xiangtan University, Xiangtan, Hunan 411105, People’s Republic of China;\\}


\begin{abstract}
Based on the modern understanding of MHD turbulence theory, we propose a new method for measuring the spectral properties of magnetic turbulence by synchrotron polarization gradient analysis. Using synthetic polarization observational data, we first confirm the feasibility of the gradient technique to determine the scaling properties of magnetic turbulence. We then apply this technique to the Galactic plane survey data from the Australia Telescope Compact Array (ATCA) and obtain the power-law spectral feature of the Galactic magnetic turbulence of $E\propto k^{-10/3}$. Gradient techniques open up a new way to obtain spectral properties of magnetic turbulence from polarization observations, such as the Low-Frequency Array for Radio Astronomy (LOFAR), in the presence of low-frequency and strong Faraday depolarization.
\end{abstract}


\section{Introduction}  \label{Intro}
Magnetohydrodynamic (MHD) turbulence plays a vital role in many key astrophysical environments  (e.g., \citealt{Beresnyak2019, Lazarian2020}). The turbulence cascade process contains several properties, such as source and sink, energy cascade efficiency, scaling behavior, magnetic field strength, eddy anisotropy, and spatial structure distribution. Exploring these properties lies in the knowledge that we can understand complex astrophysical processes and improve the development of MHD turbulence theory. With applications to spectral line data and synchrotron radiation data, one has developed various techniques to measure the nature of MHD turbulence. By analyzing position-position-velocity (PPV) spectral data, both velocity channel analysis (VCA) and velocity correlation spectrum (VCS) methods employ spectral Doppler-shifted lines to study velocity turbulence (\citealt{Lazarian2000}; see also \citealt{Lazarian2009} for a review). In analogy to the VCA and VCS techniques, using position-position frequency (PPF) data, polarization spatial analysis (PSA) and
polarization frequency analysis (PFA) techniques employ synchrotron polarization fluctuations to study magnetic turbulence (\citealt{Lazarian2016}; see also \citealt{Zhang2022} for a review).  

One of the key characteristics of MHD turbulence is its spectral distribution. When obtaining the velocity spectrum from the HI region by the VCS and VCA methods, one cannot often get self-consistent results due to the coupling between velocity and density (\citealt{Lazarian2009}), so the effects of density information need to be separated from velocity (see \citealt{Yuen2021} for separating velocity and density components). In general, the magnetic field spectrum from the HII region can be revealed by the power spectrum of synchrotron intensity and polarization intensity. Note that using synchrotron intensity can only reveal the properties of the perpendicular components of the magnetic field, which is similar to what polarization intensity can do in the high-frequency regime (without Faraday depolarization). When considering Faraday rotation of polarization emission, we can explore not only the information of the perpendicular component of the magnetic field but also the characteristics of the component parallel to the line of sight (LOS). However, due to the effect of Faraday depolarization, the power spectrum of polarization intensity will be distorted with a shortened measurable inertial (power law) range (\citealt{Lee2016, Zhang2018}), and the cascading properties of the 3D magnetic field cannot be correctly understood by the power spectrum of polarization intensity. 

To improve the measurement accuracy of the scaling index of magnetic turbulence, we attempted to extend the measurable inertial range by using the $Q$-$U$ cross intensity and the $Q$-$U$ cross-correlation intensity (\citealt{ZhangXW2023}). However, we found that these methods can only measure the power spectrum of the 3D magnetic field in the medium and high-frequency range. Given that synchrotron radiation comes from the interaction of relativistic electrons and magnetic fields, \cite{Lazarian2016} analytically established the correlation between the variance of polarization intensity and frequency to recover the scaling index of MHD turbulence (called one-point statistics in the framework of the PFA technique), confirmed by almost contemporaneous numerical work (\citealt{Zhang2016}). Note that the limitation of the polarization variance technique is that it can only measure the scaling index of a turbulent magnetic field under the condition that the  LOS mean magnetic field dominates Faraday depolarization. The advantage of polarization variance is that it provides an alternative way to measure the magnetization of the ISM (see \citealt{Guo2024} for details).

Based on a modern understanding of MHD turbulence theory (\citealt{Goldreich1995}, hereafter GS95; \citealt{Cho2002}), \cite{Lazarian2018} proposed synchrotron polarization technique to recover the properties of MHD turbulence in the framework of the PSA technique (see \citealt{Lazarian2017}) for synchrotron intensity gradient; see also \citealt{Gaensler2011} for another prototype of the gradient analysis). Later, the polarization gradient technique was successfully tested in \cite{Zhang2019a} and \cite{Zhang2020} and then used to measure magnetic field directions (\citealt{Zhang2019b, Wang2021}), magnetization (\citealt{Carmo2020}), and 3D magnetic field structure (\citealt{Hu2024}). A most significant advantage of polarization gradient techniques is that they can work well at low frequency and strong Faraday rotation region (\citealt{Zhang2019a}).

Motivated by the excellent performance of polarization gradient techniques in measuring the direction of magnetic fields, we explore whether polarization gradient techniques can determine the scaling slope of magnetic turbulence. This paper is structured as follows. Section \ref{Theor} includes a description of the theoretical basis of gradient statistics, and Section \ref{SynObs} describes how to obtain synthetical observations. Our numerical results are provided in Section \ref{Res}, following application to realistic observations in Section \ref{App}. Discussion and conclusions are positioned in Section \ref{DisCon}.

\section{Theoretical Basis of Gradient Statistics}  \label{Theor}
MHD turbulence is a complex phenomenon that governs the dynamics of magnetized plasmas in astrophysical environments. The modern understanding of MHD turbulence follows the pioneering work of GS95, the foundational framework of which emphasizes the critical balance between the Alfv\'enic wave propagation timescales and the eddy turnover times, i.e., $v_{\perp} l_{\perp}^{-1} = V_{\rm A} l_{\parallel}^{-1}$, where $l_{\perp}$ and $l_{\parallel}$ are the perpendicular and parallel scales of the eddy, respectively. Here, $v_{\perp}$ is the eddy velocity at the scale of $l_{\perp}$, and $V_{\rm A}=\langle B\rangle/\sqrt{4\pi \rho}$ the Alfv\'en velocity with the average magnetic field $\langle B\rangle$ and the density $\rho$. One of the main findings of GS95 is the introduction of scale-dependent anisotropy of $l_{\parallel} \propto l_{\perp}^{2/3}$ for incompressible turbulence at $M_A = V_{\rm L} /V_{\rm A}\simeq 1$, where $V_{\rm L}$ is the turbulence injection velocity at the injection scale $L_{\rm inj}$. This relationship arises from the Kolmogorov-like velocity scaling $v_{\perp} \propto (\epsilon l_{\perp})^{1/3}$ in the perpendicular direction ($\epsilon$ is the constant energy transfer rate in the inertial range) and indicates that turbulent eddies become increasingly elongated along the local magnetic field direction at smaller scales. The resulting energy spectrum in the perpendicular direction follows the Kolmogorov-like energy spectrum of $E(k_{\perp}) \propto k_{\perp}^{-5/3}$, where $k_{\perp}$ represents the wave-vector component perpendicular to the magnetic field. Given that the turbulent velocity of eddies obeys the relation of $v_\perp \propto l_\perp^{1/3}$ (the Kolmogorov-type cascade perpendicular to the local magnetic field), the gradients of velocity and magnetic fields can be written as 
\begin{equation}
\nabla v_\perp \propto \nabla (\delta B) \propto l_\perp^{1/3}/l_\perp =l_\perp^{-2/3}, 
\end{equation}
indicating that the gradient signal dominates at the smallest scales. According to $(\nabla v_\perp)^2 \propto (\nabla (\delta B))^2 = k_\perp E(k_\perp)$, we have a gradient spectral distribution of $E(k_\perp) \propto k_\perp^{1/3}$.

Later, GS95 theory was generalized to both $M_{\rm A}>1$ and $M_{\rm A}<1$ for compressible turbulence (\citealt{Lazarian1999}, hereafter LV99; \citealt{Lazarian2006}). How to characterize the transition between weak and strong turbulence regimes is central to understanding MHD turbulence dynamics. In the case of sub-Alfv\'enic turbulence ($M_{\rm A} < 1$), the energy cascade is dominated by weak interactions between Alfv\'en waves at large scales (called weak turbulence), with the perturbations of magnetic field perpendicular to the local magnetic field direction (LV99; \citealt{Galtier2000}). Here, we expect that the directions of magnetic field gradients are perpendicular to the local magnetic field. As the scale decreases, the turbulence transitions to strong turbulence at the scale $l_{\text{sub}} = L_{\text{inj}} M_{\rm A}^2$. 
In the range of $l_{\rm sub}$ to the dissipation scale of $l_{\rm dis}$, we have a modified scale-dependent anisotropy of $l_{\parallel} \approx L_{\text{inj}}^{1/3} l_{\perp}^{2/3} M_A^{-4/3}$,
with the relation of $v_\perp \propto l_\perp^{1/3} M_{\rm A}^{1/3}$, from which we expect also a gradient spectral distribution of $E(k_\perp) \propto k_\perp^{1/3}$.

In contrast, super-Alfv\'enic turbulence ($M_{\rm A} > 1$) behaves hydrodynamically at large scales, with isotropic eddies following the Kolmogorov spectrum of $E(k) \propto k^{-5/3}$, where the directions of gradients of magnetic field are not correlated with magnetic field direction. At the transition scale $l_{\rm sup} = L_{\text{inj}} M_{\rm A}^{-3}$, the turbulence enters to a strong turbulence regime with the characteristics of GS95, due to magnetic fields becoming dynamical importance. 

Note that the energy cascade in weak turbulence differs markedly from that in the strong regime. A weak turbulence, governed by Alfv\'en wave interactions, leads to velocity scaling of $v_\perp \propto V_L (l_{\perp}/L_{\text{inj}})^{1/2}$. With $k_\perp E(k_\perp)\sim v_\perp^2$, we have a steep velocity spectrum of $E(k_{\perp}) \propto k_{\perp}^{-2}$, reflecting the reduced nonlinear coupling. Considering $(\nabla v_\perp)^2 \propto (\nabla (\delta B))^2 = k_\perp E(k_\perp)$, we expect the gradient spectrum to present a plateau distribution. In contrast, strong turbulence adheres to the critical balance condition, from which we obtain the scaling of $\nabla v_{\perp} \propto l_{\perp}^{-2/3}$ due to the cascade dominated by the perpendicular fluctuations, resulting in the gradient spectrum of $E(k_\perp) \propto k_\perp^{1/3}$. 

The scale-dependent anisotropy of the MHD turbulence related to the transition scales is critical to understanding the interactions of turbulence in different astrophysical environments. For instance, in the sub-Alf\'enic molecular clouds, turbulence below $l_{\text{sub}}$ governs filamentary structures and star formation efficiency (e.g., \citealt{McKee2007}). In the super-Alfv\'enic galaxy clusters, the transition scale affects the heat conduction and the propagation of cosmic rays (\citealt{Schlickeiser:2002, Yan2008}). In addition, the transition scale of magnetized ISM determines the resolution at which magnetic fields can imprint observable anisotropies on continuum synchrotron emission and emission lines. Meanwhile, scale-dependent relations underpin the velocity and synchrotron emission gradient techniques (\citealt{Gonzalez2017, Lazarian2018}). 

\section{Generation of Synthetic Observations}  \label{SynObs}

\begin{table*}
  \begin{center}
\setlength{\tabcolsep}{2.2mm}
    \begin{tabular}{cccccccc} 
\hline
\text{Data} & \text{$B_{0}$} & \text{$P_0$} & \text{$M_{\rm A}$}& \text{$M_{\rm s}$} & \text{$\beta$} & \text{$\delta B_{\rm rms}/\langle B \rangle$}& Descriptions\\
    \hline
D1   & 1.0   & 2.0 & 0.65 & 0.48 & 3.67 & 0.54& sub-Alfv\'enic  \& subsonic \\
D2   & 1.0   & 0.05 & 0.58& 3.16 & 0.07 & 0.46& sub-Alfv\'enic  \& supersonic\\
D3   & 0.1   & 2.0 & 1.72&  0.45 & 29.22 & 1.12& super-Alfv\'enic  \& subsonic \\
D4   & 0.1   & 0.05 & 1.69& 3.11 & 0.59 & 1.08& super-Alfv\'enic  \& supersonic \\  
\hline 
\end{tabular}
\end{center}
\caption{MHD turbulence data. $B_{0}$ -- magnetic field strength; $P_{0}$ -- gas pressure; $M_{\rm A}$ -- Alfv{\'e}nic Mach number; $M_{\rm s}$ -- sonic Mach number; $\beta$ -- plasma parameter; $\delta B_{\rm rms}$ -- root mean square of the random magnetic field; $\langle B \rangle$ -- mean magnetic field strength. Note that $B_{0}$ and $P_{0}$ are initial parameter settings, while the other derived values are from the final snapshot data.
} \label{Tab_data}
\end{table*}

For simplicity, we assume that the non-thermal relativistic electrons keep a power-law distribution of $N(\varepsilon) d\varepsilon \propto \varepsilon^{-p}d\varepsilon$ with the spectral index $p$ ($p=2.5$ fixed in this work) and the energy $\varepsilon$. Under the assumption of negligible Faraday rotation effect, one can obtain the observable Stokes parameters from linear polarization synchrotron radiation by the following equations (see also \citealt{Waelkens2009})
\begin{equation}
I(\boldsymbol{R}) = \int_0^L dz \left(B_{\mathrm{x}}^{2}(\boldsymbol{R}) + B_{\mathrm{y}}^{2}(\boldsymbol{R})\right)^{\frac{p - 3}{4}} \left(B_{\mathrm{x}}^{2}(\boldsymbol{R}) + B_{\mathrm{y}}^{2}(\boldsymbol{R})\right),
\end{equation}
\begin{equation}
Q_0(\boldsymbol{R}) = \int_0^L dz \left(B_{\mathrm{x}}^{2}(\boldsymbol{R}) + B_{\mathrm{y}}^{2}(\boldsymbol{R})\right)^{\frac{p - 3}{4}} \left(B_{\mathrm{x}}^{2}(\boldsymbol{R}) - B_{\mathrm{y}}^{2}(\boldsymbol{R})\right),
\end{equation}
\begin{equation}
U_0(\boldsymbol{R}) = \int_0^L dz \left(B_{\mathrm{x}}^{2}(\boldsymbol{R}) + B_{\mathrm{y}}^{2}(\boldsymbol{R})\right)^{\frac{p - 3}{4}} \left(2B_{\mathrm{x}}(\boldsymbol{R})B_{\mathrm{y}}(\boldsymbol{R})\right),
\end{equation}
where $L$ represents the integration depth along the LOS and $\boldsymbol{R} = (x, y)$ corresponds to the spatial coordinates on the plane of the sky. 

When involving a Faraday rotation, we can define the complex polarization intensity as 
\begin{equation}
P(\boldsymbol{R}, \lambda^2) = Q(\boldsymbol{R}, \lambda^2) + iU(\boldsymbol{R}, \lambda^2)= \int_0^L dz \, \mathscr{P}(\boldsymbol{R}, z) e^{2i \lambda^2 R_{\rm m}(\boldsymbol{R}, z)}.
\end{equation}
Here, $\mathscr{P}$ represents an intrinsic polarization intensity density, and the exponential factor accounts for the Faraday rotation from the source at $z$ along the LOS to the observer. The Faraday rotation measure $R_{\rm m}$ is given by
\begin{equation}
R_{\rm m}(\boldsymbol{R}, z) = 0.81\int_0^z n_{\rm e}(z) B_{\rm z}(z) \, dz\ {\rm rad\ m^{-2}}, \label{eq_RM}
\end{equation}
where $n_e$ represents the thermal electron number density in units of $\rm cm^{-3}$, $B_z$ in $\mu G$, and $dz$ in pc. After considering the Faraday rotation effect, the polarization angle of synchrotron radiation is $\theta = \theta_0 + \lambda^2 R_{\rm m}(\mathbf{X}, z)$, where $\theta_0$ is the intrinsic polarization angle. Therefore, we have the modified Stokes parameters rewritten as
\begin{equation}
Q(\boldsymbol{R}, \lambda^2) = Q_0 \cos 2\theta + U_0 \sin 2\theta, \ and
\end{equation}
\begin{equation}
U(\boldsymbol{R}, \lambda^2) = U_0 \cos 2\theta - Q_0 \sin 2\theta,
\end{equation}
from which we obtain the synchrotron polarization intensity of 
\begin{equation}
P = \sqrt{Q^2 + U^2}.
\end{equation}

Considering a spatial gradient of $P=Q+iU$ on the plane of the sky, i.e., $\nabla P= \nabla Q + i\nabla U$, we have the rotationally and translationally invariant quantity (see also \citealt{Herron2018} for other more advanced diagnostics) 
\begin{equation}
|\nabla P| =\sqrt{ \left( \frac{\partial Q}{\partial x} \right)^2 + \left( \frac{\partial U}{\partial x} \right)^2 + \left( \frac{\partial Q}{\partial y} \right)^2 +\left( \frac{\partial U}{\partial y} \right)^2}, \label{eq_nablaP}
\end{equation}
in the $Q$–$U$ plane, called the magnitude of the polarization gradients. 

In this work, we use data cubes of the diffuse ISM simulated by the second-order accurate hybrid non-oscillatory code (\citealt{Cho2003}). Data cubes listed in Table \ref{Tab_data} are generated by setting the injection wavenumber of $k\simeq 2.5$, the initial magnetic field $B_0$ (along the $x$-axis direction), and the gas pressure $P_0$. The total magnetic field is $B=B_0 +\delta B$, where $\delta B$ indicates the fluctuation (random) component. Based on the output 3D data cubes including magnetic field, velocity, and density, we obtain key turbulence parameters such as the Alfv\'enic Mach number $M_{\rm A} = \langle |\textbf{v}| / V_{\rm A}\rangle$ and the sonic Mach number $M_{\rm s} = \langle |\textbf{v}| / c_{\rm s} \rangle $, to characterize the properties of MHD turbulence, where $c_s = \sqrt{P_0 / \rho}$ represents the sound speed. To generate synthetic observations, we dimensionalize the relevant physical parameters --- the magnetic field $B \simeq 3.0 \, \mu\mathrm{G}$, the thermal electron density $n_e \simeq 0.01 \, \mathrm{cm}^{-3}$, and the box size $L = 250$ pc, corresponding to the Galactic ISM environment.

\section{Numerical Results}  \label{Res}

\subsection{Images of Polarization Intensity and Gradient}

\begin{figure*}
\centering
    \includegraphics[width=0.99\textwidth,height=0.45\textheight]
    {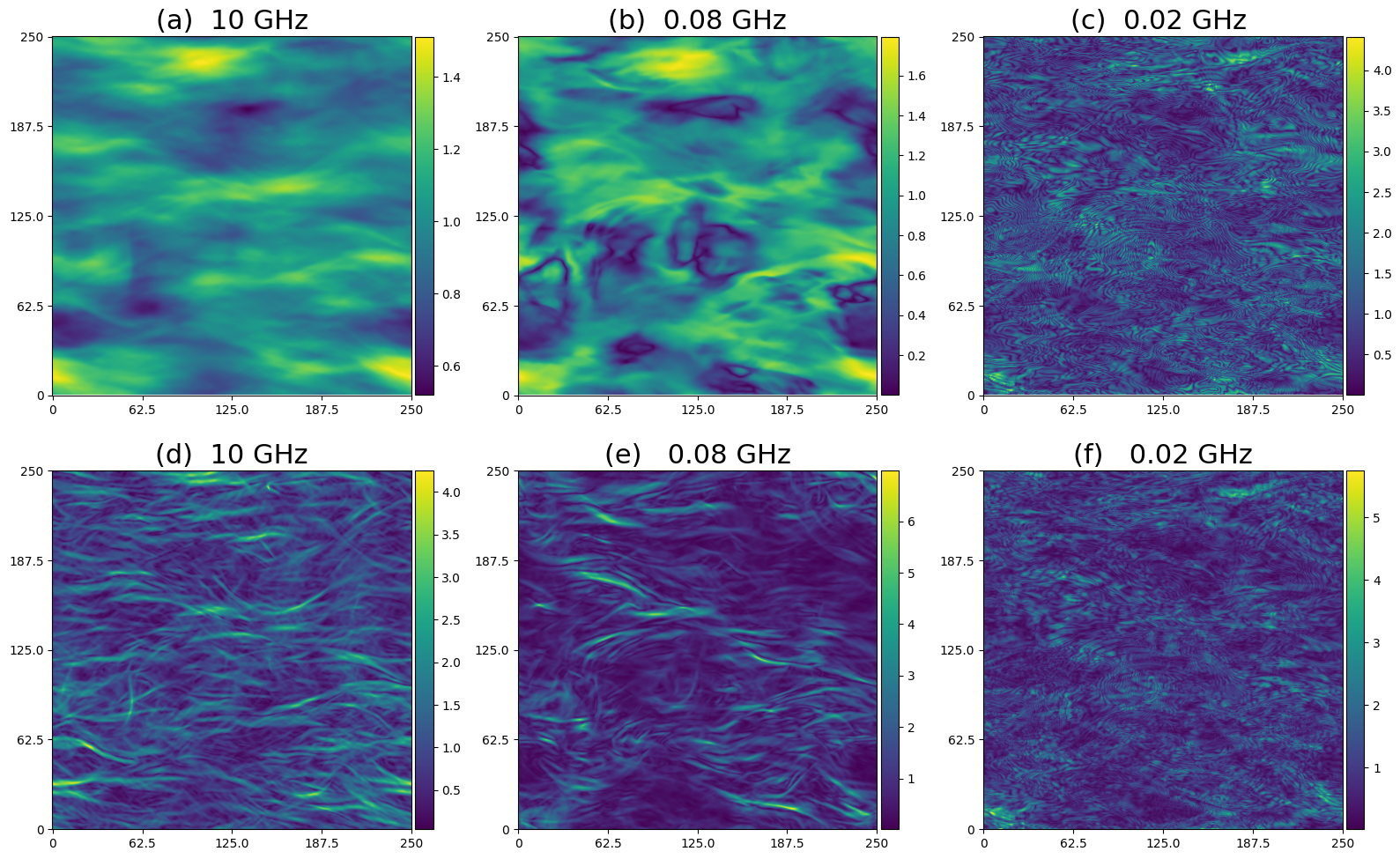}
\caption{Images of synchrotron polarization intensity (upper panels) and its gradient (lower panels) at the different frequencies, normalized in units of its mean value. The horizontal and vertical coordinates are in units of pc, with the line of sight along the $z$-axis direction. Calculations are based on D1 listed in Table \ref{Tab_data}.
} 
\label{fig_map}  
\end{figure*}

Based on Section \ref{SynObs}, we generate synthetic observations related to the Stokes parameters $Q$ and $U$, arising from synchrotron polarization emission. Figure \ref{fig_map} shows images of synchrotron polarization intensity and its gradient in the case of sub-Alf\'enic and subsonic turbulence. From the upper panels of this figure, we can see the elongated structures of polarization intensities in the horizontal direction. The reason is that the mean magnetic field direction is set in the same direction when performing the MHD turbulence simulation. In other words, the elongated structure from the image can visually reveal the direction of the mean magnetic field.

In the lower panels, we see that the image of polarization gradients presents more fine filament structures distributed horizontally due to the influence of the mean magnetic field. At the low frequency of 0.02 GHz, the details of the spatial distribution of polarization gradient cannot be seen clearly by the eye due to the appearance of small-scale noise-like structures. Since the gradient distribution of an image can significantly characterize the spatial distribution of the projected magnetic fields in the local regions, the proposed gradient techniques first traced to the direction of magnetic fields, as mentioned in Section \ref{Intro}.

\subsection{Statistical Moments of Polarization Intensity and Gradient}

\begin{figure*}
\centering
\includegraphics[width=0.99\textwidth,height=0.45\textheight]
{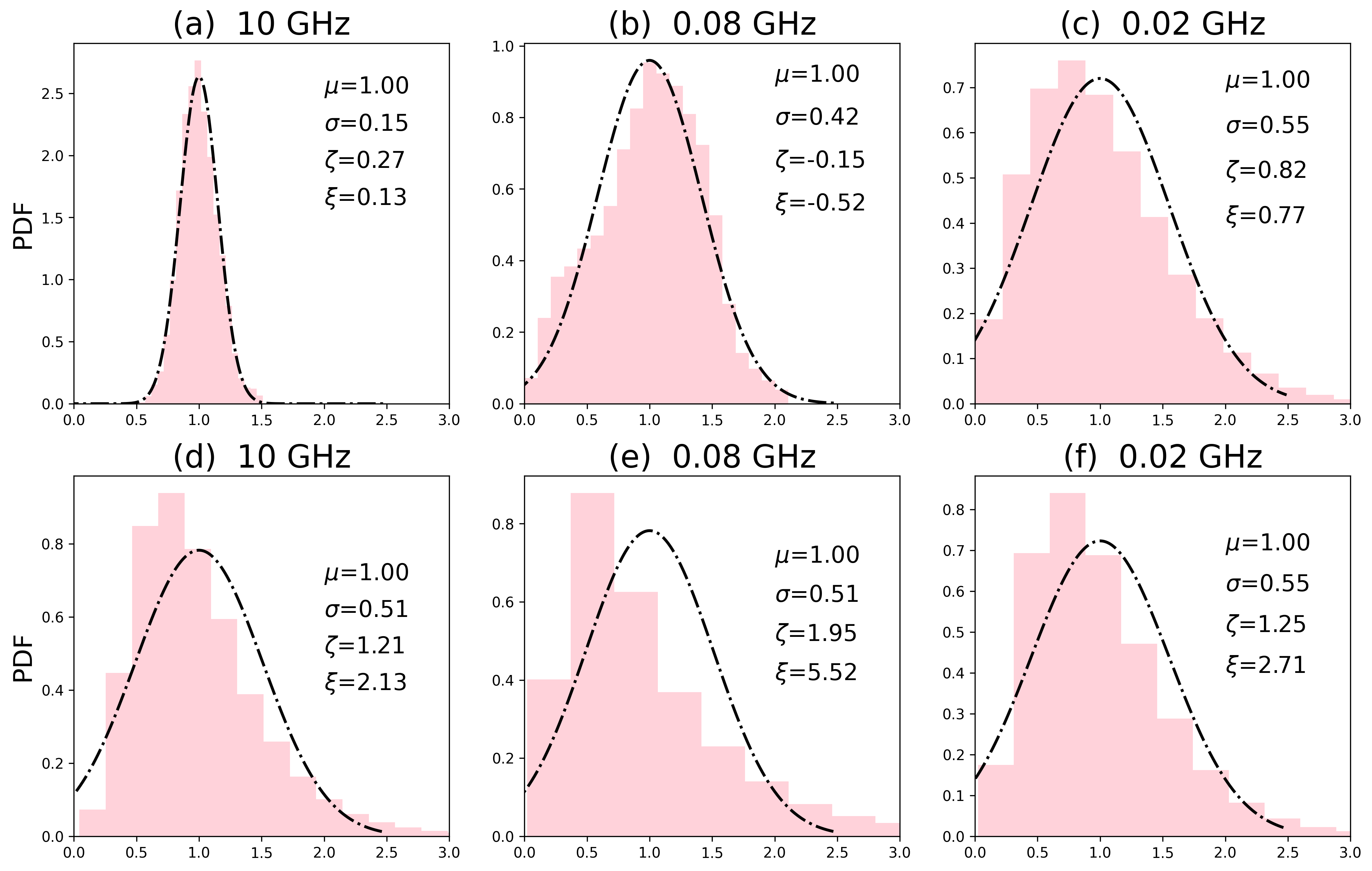}  %
\caption{Probability distribution of polarization intensity (upper panels) and its gradient (lower panels), corresponding to those maps shown in Figure \ref{fig_map}. The dash-dotted line in each panel indicates a Gaussian distribution plotted with the mean value $\mu$ and standard deviation $\sigma$. $\zeta$ and $\xi$ represent higher-order statistical measurements -- skewness and kurtosis.
} \label{fig_pdf}
\end{figure*}

The probability distribution of the polarization intensity and its gradient is plotted in Figure \ref{fig_pdf}, providing the values of the statistical moments of each order. As in Figure 1, we only consider the sub-Alf\'enic and subsonic turbulence regimes as an example. For the first and second moments --- the mean $\mu$ and standard deviation $\sigma$ of both intensity and gradient remain almost unchanged, with $\mu \simeq 1.00$ and $\sigma\simeq 0.5$, except for a smaller $\sigma = 0.15$ of the polarization distribution at a high frequency of 10 GHz (representing a good Gaussian distribution). Therefore, studies of the low-order statistical moments can hardly distinguish statistical characteristics between the polarization intensity and its gradient. For the third and fourth moments --- skewness $\zeta$ and kurtosis $\xi$ have significant differences in intensity and gradient as the frequency changes. The positive and negative values of the skewness indicate that the probability distributions have an elongated tail to the right and left, respectively. Similarly, a positive value of the kurtosis means that the probability distribution is flatter than a Gaussian distribution, and a negative value of the kurtosis implies that the distribution is more peaked than a Gaussian one. Comparing the probability distributions of the polarization intensity and its gradient, we find that with decreasing the frequency (increasing the depolarization level), gradient probability distribution exhibits consistent features different from the probability distribution of polarization intensity.

\begin{figure}
\centering
\includegraphics[width=0.45\textwidth,height=0.35\textheight]{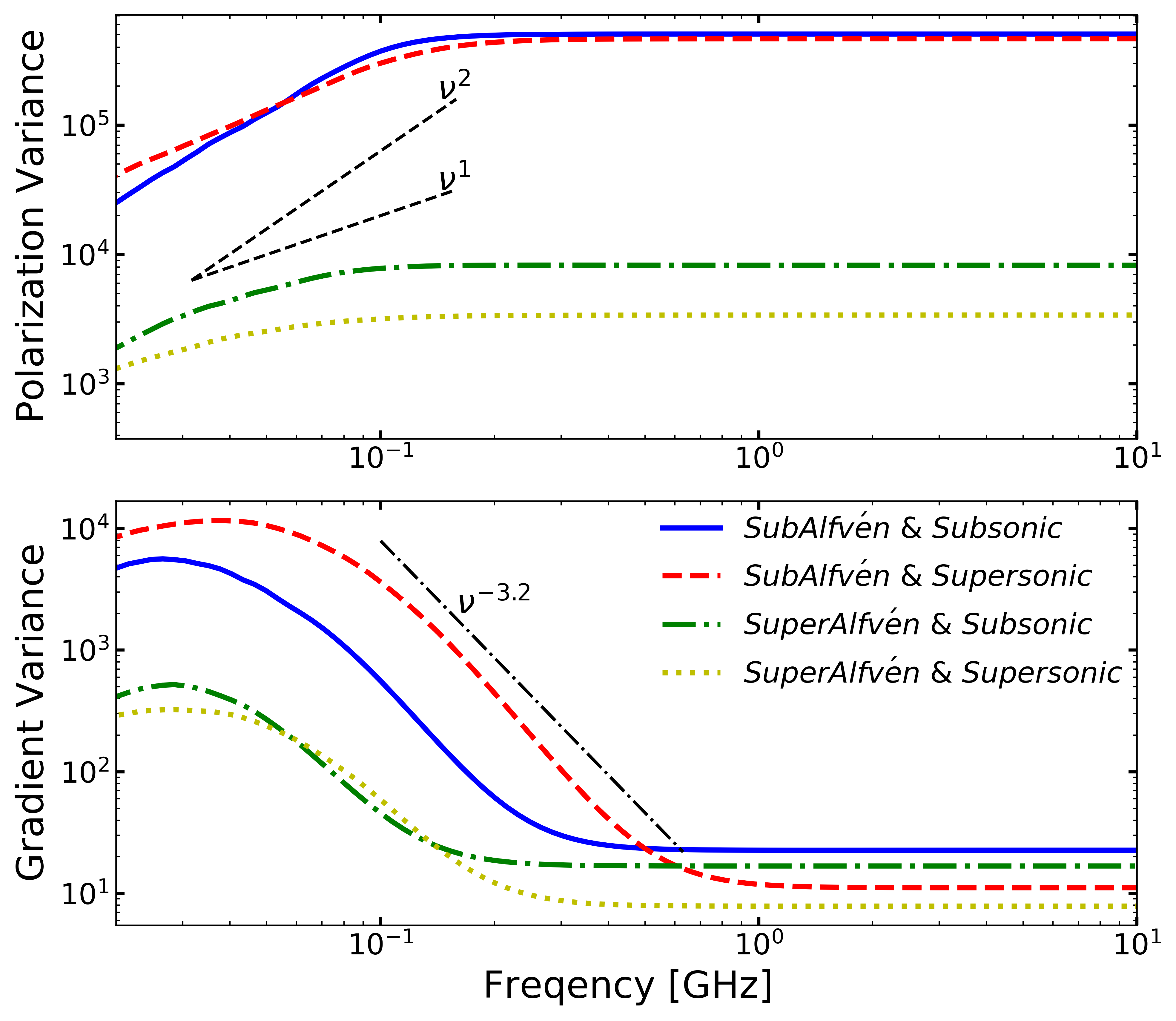} %
\caption{Variance of polarization intensity (upper panel) and its gradients (lower panel) as a function of frequency, based on D1 to D4 listed in Table \ref{Tab_data}.
} \label{fig_var}
\end{figure}

Furthermore, Figure \ref{fig_var} shows the variance of the polarization intensity and its gradient as a function of the frequency. As we can see, the intensity and gradient variances depend on the radiative frequency in the low-frequency regime and are independent of the frequency in the high frequency. The intensity variance plotted in the upper panel shows an approximate power-law relation of $\langle P^2 \rangle \propto \nu^\alpha$, with $\alpha$ in the range of 1 to 2, for four turbulence regimes explored.\footnote{To compare with the results of \cite{Zhang2016}, we rewrite the variance as $\langle P^2 \rangle \propto \nu^\alpha \propto \lambda^{-\alpha}$, where $\alpha\simeq2$ is in agreement with the case that the root mean square of the Faraday rotation density $\sigma_{\phi}$ is greater than its mean value $\bar{\phi}=\langle B_zn_e\rangle$. It is different from \cite{Zhang2016} that our current results correspond to the case of $\sigma_{\phi}\simeq \bar{\phi}$, for which we still do not have a theoretical prediction.} However, the gradient variance shows a relation of $\langle (\nabla P)^2 \rangle \propto \nu^{-3.2}$ in different regimes of turbulence. In general, the variance of the polarization intensity and its gradient exhibits different distribution characteristics, and whether the variance of the gradient can reveal the spectral index of MHD turbulence will be discussed elsewhere.

\subsection{Power Spectra of Polarization Intensity and Gradient}
\begin{figure*}
\centering
\includegraphics[width=0.96\textwidth,height=0.60\textheight]{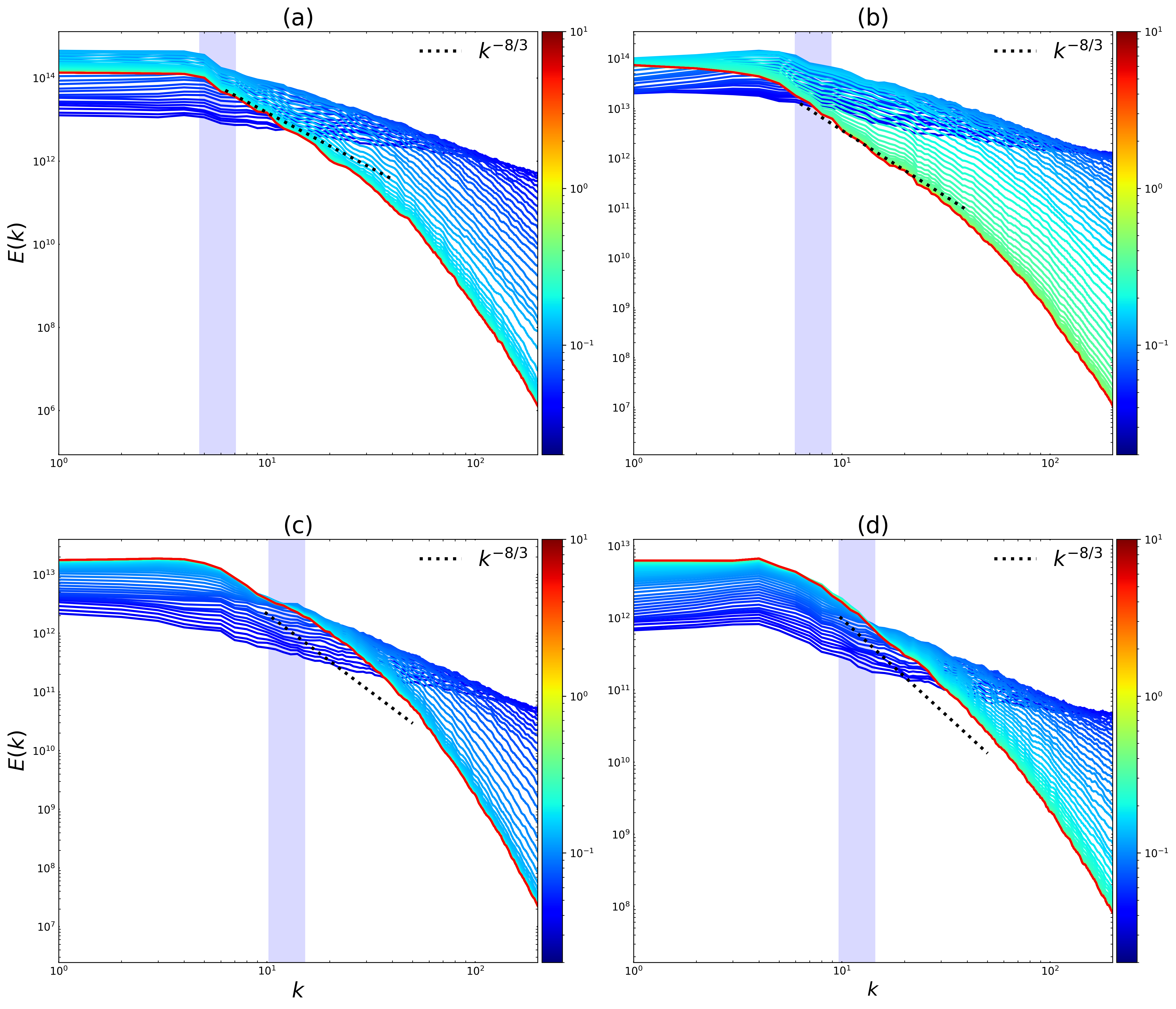}  
\caption{Power spectra of synchrotron polarization intensities calculated at individual frequencies for different turbulence regimes. The vertical light blue bands represent the transition from weak to strong turbulence. Panels (a) -- (d) correspond to D1 -- D4 listed in Table \ref{Tab_data}, respectively. The color bar plotted indicates a change in frequency in units of GHz.
} \label{fig_pol_ps}
\end{figure*}

Figure \ref{fig_pol_ps} presents the power spectra of polarization intensities in the frequency range of 0.02 to 10 GHz. As is seen in this figure, the power spectra of polarization intensities are very close to $E \propto k^{-8/3}$ at high frequencies of $\nu > 1$ GHz. This scaling index reflects the fluctuations of the perpendicular component, $B_\perp =\sqrt{B_{\rm x}^2+B_{\rm y}^2}$, of the magnetic field $\textit{\textbf{B}}$, with a 3D spectrum of $E_{\rm 3D} \propto k^{-11/3}$. As the frequency decreases, the spectrum is significantly deformed, with the amplitude of the spectrum shifting significantly upward in the large-wavenumber (small-scale) range. 
Due to the appearance of fluctuations of Faraday rotation density ($\phi=n_{\rm e}B_{\rm z}$), it leads to an increase in amplitude. However, the spectrum amplitude shifts significantly downward in the small-wavenumber (large-scale) range due to Faraday depolarization. Using the relation of $k=L\lambda^2\sigma_{\rm \phi}$ defined in \cite{Zhang2018}, we can characterize the level of Faraday depolarization that is significant at $k<L\lambda^2\sigma_{\rm \phi}$ and insignificant at $k>L\lambda^2\sigma_{\rm \phi}$. According to Equation (\ref{eq_RM}), we obtain values of the Faraday rotation measure in the range of $-0.3$ to $0.3\rm\ rad\ m^{-2}$ in the case of sub-Alfv\'enic and subsonic turbulence. In this regard, we have the maximum amount of change in the rotation angle of $\Delta \theta\simeq 0.2^\circ$ at 10 GHz and of $\Delta \theta\simeq 21\pi+85^\circ$ at 0.02 GHz. As a result, the change in the observed frequency (or wavelength) introduces a very significant Faraday depolarization, and the power spectrum of the polarization intensity in the range of low frequencies does not reveal the properties of the magnetic field.

\begin{figure*}
\centering
\includegraphics[width=0.96\textwidth,height=0.60\textheight]{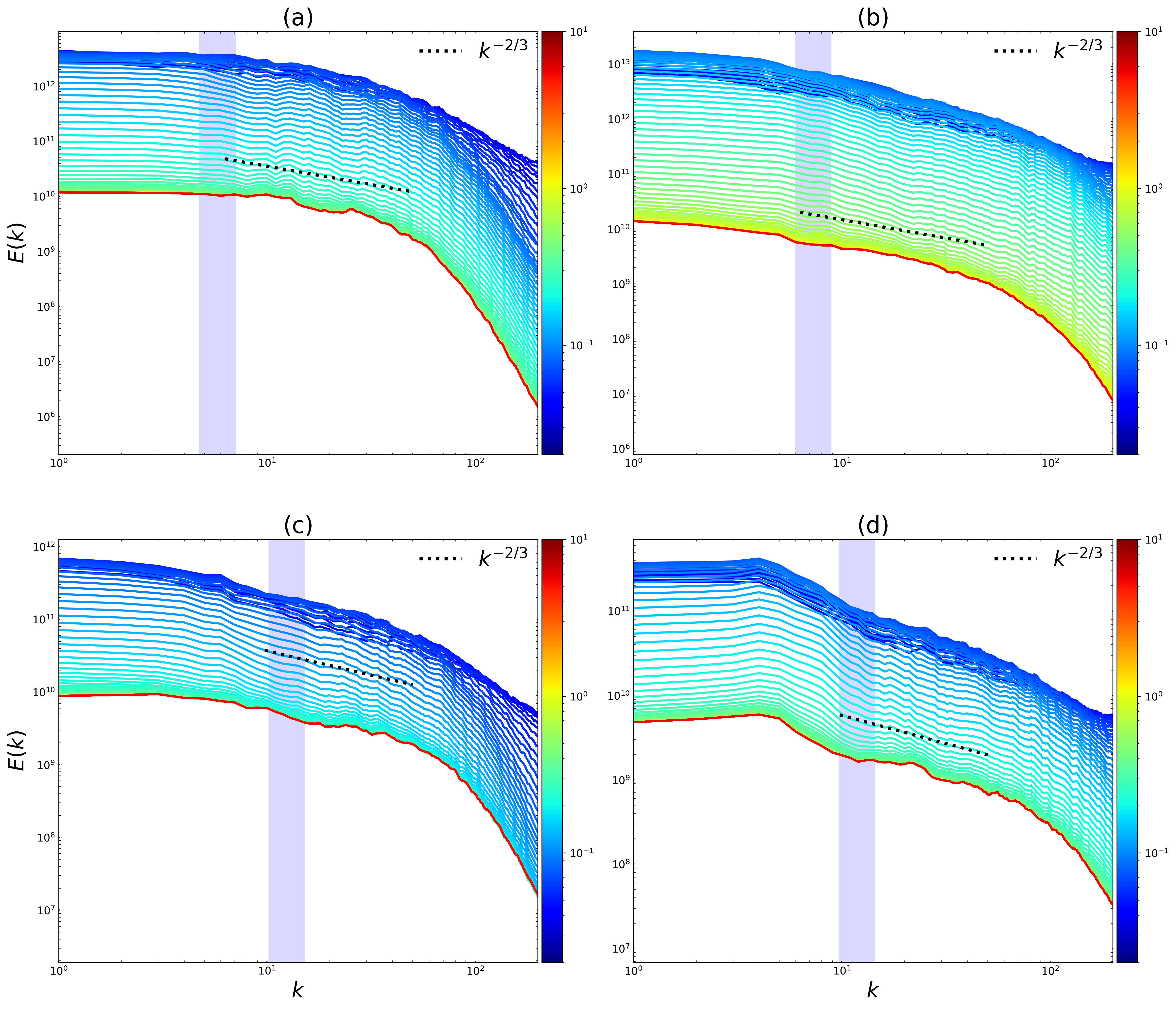}  
\caption{Power spectra of polarization gradient calculated at individual frequencies for different turbulence regimes. The vertical light blue bands represent the transition from weak to strong turbulence. Panels (a) -- (d) correspond to D1 -- D4 listed in Table \ref{Tab_data}, respectively. The color bar plotted represents a change in frequency in units of GHz.
} \label{fig_grad_ps}
\end{figure*}

In the same frequency range of 0.02 to 10 GHz, we explore the power spectrum of the polarization gradient in Figure \ref{fig_grad_ps}, from which we can see that the polarization gradients show the power spectra of $E \propto k^{-2/3}$ in the inertial range for different regimes of turbulence. With decreasing frequency, we see that the power spectra move upwards overall. Unlike the polarization intensity spectra shown in Figure \ref{fig_pol_ps}, the polarization gradient spectra are not distorted by Faraday depolarization. Interestingly, the spectra of polarization intensity in the low-frequency range cannot recover the scaling index because of Faraday depolarization, whereas the spectra of the polarization gradient still exhibit good power-law characteristics. This is because the intensity is directly related to the change in the polarization vectors, while the polarization gradient is not affected by the change in the polarization direction. 

Suppose that MHD turbulence maintains a general cascade law of $v \propto \delta B \propto k^{-m}$ ($m$ stands for the type of MHD turbulence), we have a 1D power spectrum of $E_{\rm 1D} \propto k^{-2m}/k=k^{-2m-1}$. Here, setting $m=1/3$, we have $E_{\rm 1D} \propto k^{-5/3}$ corresponding to a Kolmogorov spectrum. When considering velocity and magnetic field gradients of $\nabla v \propto \nabla (\delta B) \propto k^{-m+1}$, we have a 1D gradient power spectrum of $E_{\rm 1D} \propto k^{-2m+2}/k=k^{-2m+1}$. When setting $m=1/3$, we get the gradient spectrum of $E_{\rm 1D} \propto k^{1/3}$. As shown in Figrue \ref{fig_pol_ps}, the 2D power spectrum of polarization intensity of $E_{\rm 2D,P} \propto  k^{-2m-2} =k^{-8/3}$ reflects the underlying 3D spectral properties of the perpendicular component (with $E_{\rm 3D} \propto k^{-11/3}$) of the magnetic field at the high-frequency regime. For the power spectrum of polarization gradient plotted in Figure \ref{fig_grad_ps}, we have the power-law relation of $E_{\rm 2D, \nabla P} \propto k^{-2m+1-1} = k^{-2m}=k^{-2/3}$. Based on the above analysis, we establish the following relationship 
\begin{equation}
E_{\rm 2D, \nabla P} \propto k^{-2m} \propto  k^{2} E_{\rm 2D, P} \label{eq_PandGrad}
\end{equation}
between the 2D polarization intensity spectrum and the 2D polarization gradient one. From Equation (\ref{eq_PandGrad}), we can write a 3D spectrum of the turbulent magnetic field as
\begin{equation}
E_{\rm 3D}\propto E_{\rm 2D, P}/k\propto E_{\rm 2D, \nabla P}/k^{3}\propto k^{-2m-3}, \label{eq_2Dto3D}
\end{equation}
which can recover the power-law properties of the 3D magnetic field from the polarization gradient power spectrum. Therefore, the spectra of the polarization gradient provide a new way to recover the power-law exponent of magnetic turbulence when the polarization intensity spectrum cannot work. 

We stress that the power spectrum of the polarization gradient may include an extra numerical error from the gradient calculation when compared with the power spectrum of the polarization intensity. When obtaining the power-law properties of magnetic turbulence by gradient techniques, we thus recommend using the highest possible resolution to reduce the bias caused by numerical errors. The advantage of the polarization gradient technique is that it can obtain the magnetic field properties in the case of low frequency and strong Faraday depolarization, where the traditional polarization vector method and the power spectrum of polarization intensity cannot work.

\section{Application to Real Observations}  \label{App}

\begin{figure*}
\centering
\includegraphics[width=0.96\textwidth,height=0.40\textheight]{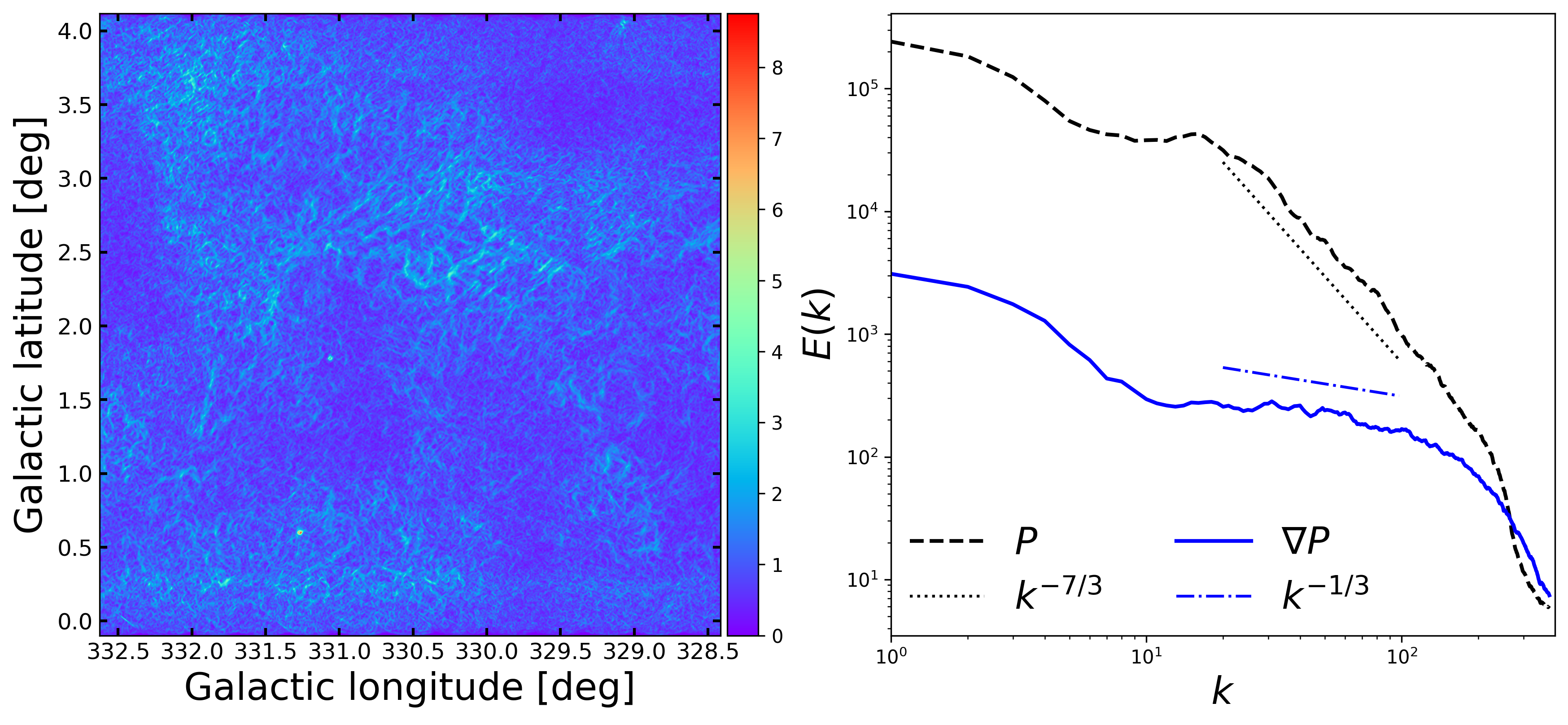}  
\caption{Application to data of the ATCA. The left panel is a gradient image of the polarization intensity, in units of gradient mean value. The right panel is the power spectra of the polarization intensity and its gradient.
} \label{fig_obs}
\end{figure*}

To test the feasibility of a polarization gradient to measure the scaling slope of the Galactic ISM, we used radio-continuum images of an 18 degree$^2$ patch of the Galactic plane, observed with the Australia Telescope Compact Array (ATCA) at a frequency of 1.4 GHz (\citealt{Gaensler2001,McClure-Griffiths2001}). The ATCA data, adopted in our previous work (\citealt{Zhang2019a,Guo2024,Wang2024}), are ideal for testing our methods and obtaining the properties of the warm ionized medium. To avoid the margin effect of the images, we extract the largest square area of $759 \times 759$ pixels from the initial image, which corresponds to a spatial area of about 18 degree$^2$. 

The results we applied to the ATCA data are shown in Figure \ref{fig_obs}, the left panel of which is an image of the gradient of polarization intensity arising from magnetized turbulence in Galactic diffuse and ionized ISM. As we can see, a complex filamentary web of the polarization gradient reveals fluctuations in the magnetic field and gas density. From the right panel, we can see that the power spectrum of the polarization gradient has a scaling slope of $E \propto k^{-1/3}$ in the inertial range, which can match the polarization spectrum with a power law of $E \propto k^{-7/3}$, according to Equation (\ref{eq_PandGrad}). The latter in the range of $k>15$ reveals the power-law nature of turbulence while the range of $k<15$ is concave downward due to the influence of depolarization. According to Equation (\ref{eq_RM}), we have a rotation angle of $\Delta \theta\simeq 46.43^\circ \lambda^2 B_{\rm z} n_{\rm e} L\simeq 12.59^\circ$ at 1.42 GHz, setting $n_{\rm e}$, $B_{\rm z}$, and $L$ to typical ISM values provided in Section \ref{SynObs} (see also \citealt{Taylor2003}). Interestingly, the gradient spectrum, the level of its concave downward attenuated, can reflect the power-law characteristic of $E_{\rm 3D} \propto k^{-10/3}$ of the ISM turbulence (derived by Equation  (\ref{eq_2Dto3D})). 

\section{Discussion and Conclusions}  \label{DisCon}
The development of the gradient technique is rooted in the anisotropy of MHD turbulence theory, as described in Section \ref{Theor}, with the eddies elongated along the direction of the local magnetic field. The spatial gradient of the synchrotron polarization intensity distributes a pronounced filament structure (see Figure \ref{fig_map}), which provides an intuitive possibility for the gradient technique to trace the direction of the magnetic field in the first place (e.g., \citealt{Lazarian2018}). Our current work advances the development and application of gradient techniques and explores the use of gradient techniques to measure the scaling properties of the MHD turbulence cascade.

The traditional approach to analyzing polarization data is called the Faraday rotation synthesis technique (\citealt{Burn1966,Brentjens2005}), which can provide information on the strength and direction of the average regular field component along the line of sight (\citealt{Sokoloff1998}). As the rotation angle is sensitive to the sign of the magnetic field direction, only large-scale regular fields give rise to Faraday rotation. In contrast, the Faraday rotation contributions from small-scale turbulent/random fields largely cancel along the line of sight (\citealt{Beck2015}). In addition, the coupling between density and magnetic fields introduces ambiguities, which means that Faraday rotation synthesis fails to disentangle magnetic field information. Compared with Faraday rotation synthesis,\footnote{Interested readers can refer to \cite{Ho2019} for comparing Faraday tomography and polarization gradients. For the former, the Faraday depth changes its sign along the line of sight, making the measured magnetic field parallel ambiguous (e.g., \citealt{Ferriere2016}).} The advantage of the polarization gradient technique is that it is not affected by Faraday depolarization and can unlock new capabilities for understanding turbulence spectrum, magnetic field structure, and eddy anisotropy. Our polarization gradient technique complements Faraday rotation synthesis by enabling turbulence studies.

Comparing the different order statistical moments of the polarization intensity and the polarization gradient, we found significant differences between them. The statistical moment of polarization intensity depends significantly on the Faraday depolarization effect, while the dependence of the gradient statistical moment on Faraday depolarization is weakened. It indicates that gradient statistics are more advantageous than polarization intensity statistics for measuring magnetic turbulence properties. The variance of polarization gradient distributes a power-law relation as a function of the radiative frequency, which provides a possibility for determining the scaling slope of MHD turbulence, similar to the polarization variance done in \cite{Zhang2016}. 

We stress that the measurement of the power spectrum of the polarization gradient is established on the anisotropy of MHD turbulence theory. The correlation of the observable polarization signal is limited by the correlation scale of the strong turbulence, e.g., $l_{\rm sub}$ and $l_{\rm sup}$ for sub- and sup-Alfv\'enic regimes, respectively. Therefore, a large-scale regular magnetic field cannot interfere with applying gradient techniques under conditions within the range of strong turbulence. The gradient techniques provide insights into turbulent magnetic turbulence mixing large-scale regular fields, making them a promising tool for studying magnetized plasmas in diverse astrophysical environments. When we generated isotropic 3D turbulence data (see Section 3.1 of \citealt{Zhang2016} for details) to perform polarization gradient studies, we found that the gradient statistics do not reflect the expected relationship provided in Equation (\ref{eq_PandGrad}). In addition, the statistics of the gradient structure function of the polarization intensity can also reflect the power-law nature of magnetic turbulence. However, due to the high numerical resolution required for the statistics of structure functions (see the Appendix section of \citealt{Lee2016}), we did not study the structure functions of gradients in this paper.

For simplification, we only consider a case of the spatially coincident synchrotron emission and Faraday rotation regions, arising from the diffusive interstellar medium in the Milky Way. Namely, the rotation measure is only attributed to the emitting source region but not any intervening material between the observer and the emitting source region along the line of sight. For spatially separated synchrotron emission and Faraday rotation regions, we found that the polarization gradient technique could also successfully trace the projected magnetic field independent of Faraday depolarization (\citealt{Wang2021}). With the enhanced resolution of the polarization data available, it is possible to apply our techniques to extragalactic sources, e.g., to nearby galaxies (see \citealt{Liu2023} for an example) and radio lobes of radio galaxies. In this regard, the polarization radiation and Faraday rotation of the Milky Way acts as a foreground. When the angular extent of the radio source is relatively small, the influence of the foreground amounts to essentially invariance along both lines of sight Faraday rotation and polarization intensity. Therefore, those factors should not interfere with the gradient analysis.

In conclusion, we confirmed the possibility of the polarization gradient technique to determine the power-law properties of MHD turbulence and applied it to the ATCA observations, with the finding of the scaling slope of $E\propto k^{-10/3}$. Our current work provides a new way to reveal the power-law properties of magnetized turbulence from low-frequency polarization information, such as data from low-frequency arrays for radio astronomy.

\begin{acknowledgments}
We thank the anonymous referee for valuable comments that significantly improved the quality of the paper. We thank the support from the National Natural Science Foundation of China (grant No. 12473046) and the Hunan Natural Science Foundation for Distinguished Young Scholars (No. 2023JJ10039). 
\end{acknowledgments}
\vspace{5mm}

          
\bibliography{ms}{}
\bibliographystyle{aasjournal}

\end{document}